\documentclass[twocolumn,showpacs,preprintnumbers,amsmath,amssymb,footinbib,prl]{revtex4}

\usepackage{epsfig}
\usepackage{graphicx}
\usepackage{color}

\def\unit#1{\ensuremath{\mathrm{~#1}}}

\def\braket#1{\ensuremath{\left<#1\right>}}
\def\TBKT{\ensuremath{T_{\mathrm{BKT}}}}
\def\nBKT{\ensuremath{n_{\mathrm{BKT}}}}
\def\poub#1{}

\begin{document}

\title{Observation of a 2D Bose-gas: from thermal to quasi-condensate to superfluid}

\date{\today}

\author{P.~Clad\'e}
\altaffiliation[Permanent Address: ]{Laboratoire Kastler Brossel, UPMC, CNRS, ENS, Paris, France}
\affiliation{
Atomic Physics Division, National Institute of Standards and Technology, Gaithersburg, Maryland 20899-8424, USA}
\affiliation{Joint Quantum Institute, NIST and University of Maryland, College Park, Maryland 20742, USA}

\author{C.~Ryu}
\altaffiliation[Present Address: ]{Quantum Institute, LANL, Los Alamos, NM }
\affiliation{
Atomic Physics Division, National Institute of Standards and Technology, Gaithersburg, Maryland 20899-8424, USA}
\affiliation{Joint Quantum Institute, NIST and University of Maryland, College Park, Maryland 20742, USA}

\author{A.~Ramanathan}
\affiliation{
Atomic Physics Division, National Institute of Standards and Technology, Gaithersburg, Maryland 20899-8424, USA}
\affiliation{Joint Quantum Institute, NIST and University of Maryland, College Park, Maryland 20742, USA}

\author{K.~Helmerson}
\affiliation{
Atomic Physics Division, National Institute of Standards and Technology, Gaithersburg, Maryland 20899-8424, USA}
\affiliation{Joint Quantum Institute, NIST and University of Maryland, College Park, Maryland 20742, USA}

\author{W.D.~Phillips}
\affiliation{
Atomic Physics Division, National Institute of Standards and Technology, Gaithersburg, Maryland 20899-8424, USA}
\affiliation{Joint Quantum Institute, NIST and University of Maryland, College Park, Maryland 20742, USA}

\begin{abstract}
We present experimental results on a Bose gas in a quasi-2D geometry near the Berezinskii, Kosterlitz and Thouless (BKT) transition temperature. By measuring the density profile, \textit{in situ} and after time of flight, and the coherence length, we identify different states of the gas. In particular, we observe that the gas develops a bimodal distribution without long range order. In this state, the gas  presents a longer coherence length than the thermal cloud; it is quasi-condensed but is not superfluid. Experimental evidence indicates that we observe the superfluid transition (BKT transition). 
\end{abstract}

\maketitle
\def\th{\mathrm{th}}
One of the most fascinating aspects of a Bose gas in the degenerate regime is the role of dimensionality. A 2D interacting Bose gas is superfluid at low enough temperature\cite{Kosterlitz1973,Berezinskii1972}. However, by contrast to the 3D case, there is no long range coherence and the coherence decays as a power law\cite{Kosterlitz1973,Berezinskii1972}. At temperatures above the Berezinskii-Kosterlitz-Thouless (BKT) transition temperature, the gas is not superfluid. Due to proliferation of free vortices, the quasi-condensate (QC) is fractured into small regions of nearly uniform phase, whose size, which corresponds to the typical length of the exponential decay of the coherence, is larger than the thermal de~Broglie wavelength ($\lambda=\sqrt{2\pi\hbar^2/Mk_{B}T}$, where $T$ is the temperature and $M$ the atomic mass). For higher T, this size becomes smaller and approaches $\lambda$, as the gas crosses over to the thermal phase. 

Experiments on 2D bosonic systems, such as two-dimensional $^4\mathrm{He}$ films \cite{Bishop1978} and trapped Bose gases \cite{Hadzibabic2006,Kruger2007}, are able to show signatures of the BKT transition. Other systems, such as the superconducting transition in arrays of Josephson junctions \cite{Resnick1981} and a two dimensional lattice of (3D) Bose-Einstein condensates \cite{Schweikhard2007}, also exhibit a similar transition. Another interesting observation was in two dimensional spin polarized atomic hydrogen on liquid $^4$He \cite{Safanov1998} where a reduction in three-body dipolar recombination (which is usually associated with condensation) was observed well above the BKT transition temperature. This observation results from a reduction of density fluctuations, which corresponds to quasi-condensation \cite{Kagan2000} \cite{footnote1}. 

In this letter, we present evidence of transitions in a quasi-2D Bose gas from thermal (normal gas), to quasi-condensate without superfluidity, to superfluid quasi-condensate (BKT transition). We explicitly identify the theoretically expected non-superfluid quasi-condensate, a feature not clearly seen in other experiments on a 2D trapped Bose gases \cite{Hadzibabic2006,Kruger2007}. We use an interferometric method to study the coherence of the gas down to distances smaller than the thermal de~Broglie wavelength. Our results can be understood using the local density approximation (LDA) on a model~\cite{Prokofev} of a homogeneous system.
More recently, calculations for a trapped system have been carried out using classical-quantum field methods~\cite{Bisset} and quantum Monte Carlo methods~\cite{Holzmann2007}. 


The BKT transition occurs at a universal value of the superfluid density $n_s=4/\lambda^2$. However, the total density at the transition, $\nBKT$, is not universal and depends on the strength of the interactions~\cite{Kosterlitz1973}. Interactions in a weakly interacting Bose gas trapped in a quasi-2D geometry (trap with tight confinement, at large frequency $\omega_0$, along one axis) are described by a dimensionless coupling constant $\tilde g=\sqrt{8\pi} a/l_0$~\cite{Petrov2004}, where $a$ is the 3D scattering length and $l_0 = \sqrt{\hbar / M\omega_0}$ the extent of the ground state along the tight direction. In the limit of weak interactions ($\ln(1/\tilde g) < 1$), $\nBKT\lambda^2 \approx \ln(C/\tilde g)$~\cite{Fisher1988,Popov1983}. Monte-Carlo simulations that calculate the density as a function of the chemical potential give $C\approx 380$~\cite{Prokofev}. We can use those simulations for our trapped system by applying a LDA (see Ref.~\cite{Prokofev}). This is valid even near the BKT transition if the healing length defined at  $\TBKT$, $r_c \simeq 2\lambda / \sqrt{2\pi\tilde{g}}$, is smaller than the typical size of the cloud (from the chemical potential given in ~\cite{Prokofev}, we find that the Thomas-Fermi radius of the cloud scales as $R\approx \sqrt{\tilde g}R_T$, where $R_T = \sqrt{2k_BT/m\omega_\perp^2}$ is the size of the thermal cloud, $\omega_\perp$ being the frequency of the trap in the 2D plane). This implies that to satisfy the LDA and thus have a BKT transition at a higher temperature than the true Bose-Einstein condensation (BEC) transition~\cite{Petrov2000}, one must satisfy $\hbar\omega_\perp < \tilde gk_BT$. For our typical experimental parameters, $\tilde g \simeq 0.02$, $T \simeq 100\unit{nK}$ and $\omega_\perp/2\pi \simeq 20\unit{Hz}$, so the LDA is reasonable. 

\begin{figure}
	\includegraphics[width=.7\linewidth]{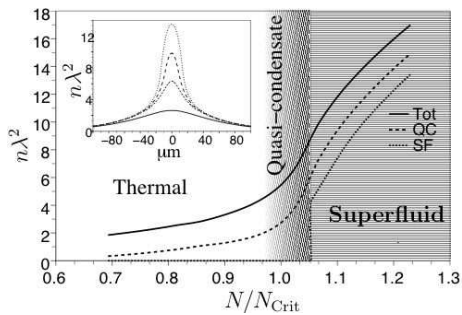}
	\caption{\label{fig:Theorie}Calculated total density (Tot), quasi-condensate (QC) density and superfluid (SF) density at the center of the cloud in a trap as a function of the total number of atoms. Inset: density profile for different numbers of atoms, from $N/N_\mathrm{Crit}\approx 0.9$ (solid line) to   $N/N_\mathrm{Crit}\approx 1.1$ (dotted line). }
\end{figure}

Applying the LDA to the calculation of Refs.~\cite{Prokofev}, we get (Fig.~\ref{fig:Theorie}) the total density at the center of the trap as a function of the total number of atoms $N$ (normalized to the BEC critical number for 2D trapped, non-interacting atoms, $N_{\mathrm{Crit}} = \pi^2/6\,(k_BT/\hbar\omega_\perp)^2$~\cite{Bagnato}). When $N$ approaches $N_{\mathrm{Crit}}$, the central density increases rapidly and the calculation predicts the appearance of a bimodal distribution (Fig.~\ref{fig:Theorie}, inset). This increase coincides with a reduction in density fluctuations, which Refs.~\cite{Prokofev} define as the appearance of a quasi-condensate. The BKT phase transition occurs when the density at the center reaches \nBKT. For $T>\TBKT$  ($n < \nBKT$), the 2D cloud will be broken up by the free vortices into multiple, phase independent, local condensates (non superfuid quasi-condensate). For $T<\TBKT$ ($n > \nBKT$), there will be a superfluid (quasi-)condensate with phase fluctuations (due to phonons and bound vortex pairs). We do not expect to see a dramatic change of the \textit{in situ} image at the phase transition, because it affects mainly the phase coherence and not the spatial density profile. However, in our experiment, phase coherence is revealed in time of flight (TOF) or by interferometry. By all these methods, we see the progression from thermal to quasi-condensate to superfluid.

Our experiment uses sodium atoms confined in a single, horizontal, 
quasi-2D, optical dipole trap, similar to that described in 
\cite{Gorlitz2001}. The infrared (1030\unit{nm}) trap laser beam, is 
spatially 
filtered through an optical fiber and collimated. The beam is then focused into a sheet of light using a cylindrical lens and projected onto the atomic cloud using a telescope. The sheet of light has a waist ($1/e^2$ radius) in the vertical direction of $\approx 9\unit{\mu m}$, and a Rayleigh range of 440\unit{\mu m}. The waist in the horizontal direction perpendicular to the propagation axis of the laser is 800\unit{\mu m}. However, due to a small distortion of the laser intensity profile in this direction, atoms see a trapping frequency higher than it would be for a Gaussian profile with that waist. This results in a trapped cloud of atoms that is approximately a circular disk with an aspect ratio of 50. For a typical value of the laser power (500\unit{mW}), the trapping frequencies are 1\unit{kHz} and 18 \unit{Hz} in the vertical and horizontal directions, respectively.  

We initially confine sodium atoms in a magnetic trap and cool them using rf-induced evaporation \cite{Kozuma1999}. Before reaching quantum degeneracy, we adiabatically transfer the atoms to the optical dipole trap. After 5~s of evaporation in the dipole trap, we reduce the depth of the trap (initially around 5\unit{\mu K}) during 1~s to a final value ranging from 5\unit{\mu K} to 2\unit{\mu K}  and then hold for 5~s to complete the evaporative cooling. We set the final temperature by choosing the final trap depth. 
We vary the phase space density keeping temperature constant (same trap depth) and controling the initial number of trapped atoms. For a trap with a vertical confinement of $\omega_0 /2\pi= 1\unit{kHz}$ the temperature is around $T = 100\unit{nK}$, which corresponds to $k_BT \approx 2\hbar\omega_0$, so the thermal motion is not completely frozen out along the tight direction. However, the quasi-condensed atoms are well confined in 2D since the mean field never exceeds 1\unit{kHz}, and therefore we expect to see 2D physics. 

\begin{figure}
	\includegraphics[width=.99\linewidth]{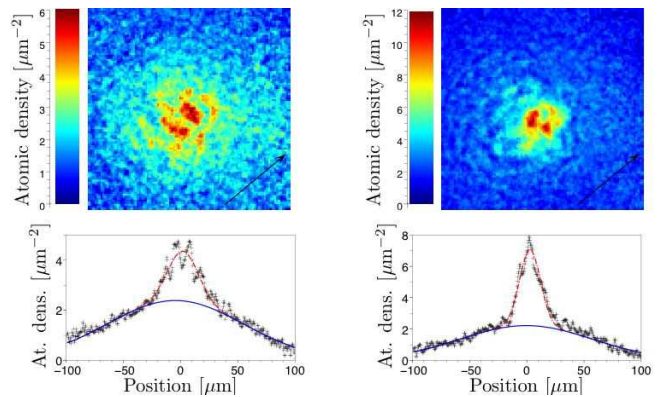}
	\caption{\label{PictAfterTOF}Density profile of the cloud after 5\unit{ms} TOF. The two images (with the same axis scale, but different color scale) are taken for two different numbers of atoms. The graph represent the cross section of the cloud along a line passing through the center of the cloud along the direction of the trap laser (arrow). The dot-dashed (red) line is a fit to the sum of two Gaussians, while the solid (blue) line is the wider of the Gaussians.}
\end{figure}

Figure~\ref{PictAfterTOF} shows two absorption images of the cloud, after 5\unit{ms} of TOF, for the same temperature but different number of atoms. Both pictures show bimodality and density fluctuations. These fluctuations, which appear only after TOF, are manifestations of the \textit{in situ} phase fluctuations of the quasi-condensate. In spite of the similarities (bimodality and density fluctuations), there are significant differences between the two pictures. The one on the left with fewer atoms is clearly broken up, while the one on the right is more compact and the narrow part of the fit (dot-dashed/red curve) is not as wide. 



\begin{figure}
	\includegraphics[width=.7\linewidth]{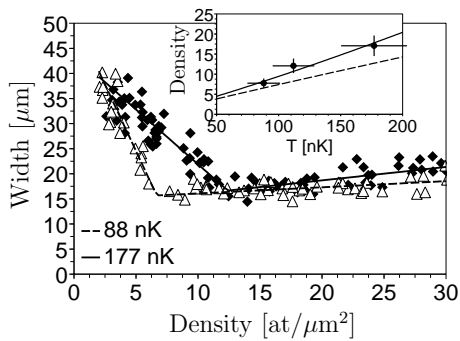}
	\caption{\label{BECWidth}Width of the narrow part of the cloud, obtained by fitting 5~ms TOF images with two Gaussians, as a function of the fitted peak 2D density, for two different temperatures. The lines are linear fits to two portions of the data, and are used to determine a transition point. Inset: 2D ``in situ" peak density~\cite{footnote2} at the transition point as a function of the @hyphenation{tem-pera-ture}. The dashed line is the theoretical value $\nBKT\lambda^2 = \ln(380/\tilde g)$ and the solid line is the same value corrected using a 3D~model~\cite{Holzmann2008bis}. We add a $\pm 15\%$ uncertainty~\cite{uncertainties} primarily due to the disagreement of the different methods used for calibration (for the temperature: TOF and interferometric measurement of $\lambda$; for the density: theoretical absorbtion cross section and \textit{in situ} measurement of the Thomas-Fermi radius).}
\end{figure}


In figure~\ref{BECWidth} we plot the width of the narrow part as a function of the peak 
density for two different temperatures (trap depths). For both temperatures, the width initially decreases with increasing density (presumably due to a decrease of free vortices) and then, beyond a certain point, increases slowly (presumably due to repulsive interactions between atoms). The central density at which the width is a minimum depends on the temperature of the gas: we identify this as the point when atoms at the center of the trap cross the BKT transition. In the inset of Fig.~\ref{BECWidth}, we plot this critical density~\cite{footnote2} as a function of the temperature. The dashed line is the theoretical prediction~\cite{Prokofev} $\nBKT\lambda^2 = \ln(380 / \tilde{g})$ and the continious line is this theoretical prediction corrected by a factor taking into acount thermal excitation of modes in the tight direction, as done in Ref.~\cite{Holzmann2008bis}. 
The position of the point of minimum width agrees with the theoretical prediction, supporting its identification  as the BKT transition in the central region of the trap. 


\begin{figure}
\noindent
\includegraphics[width=.9\linewidth]{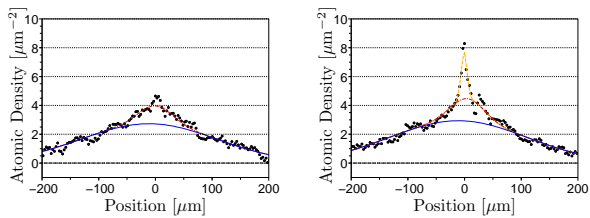}
	\caption{\label{10mstof} Cross section of the density profile of the cloud after 10\unit{ms} of TOF, fitted with the sum of three Gaussian. Left: for $n<\nBKT$ ($T>\TBKT$); right: for $n>\nBKT$ ($T<\TBKT$). The data are taken with the same trap depth as Fig~\ref{PictAfterTOF}.}
\end{figure}
With this interpretation, the picture on the left of Fig.~\ref{PictAfterTOF} is taken for $T>\TBKT$ and contains a thermal component (solid/blue curve) and a non superfluid quasi-condensate (dot-dashed/red curves) whereas the picture on the right ($T<\TBKT$) contains the thermal part and a QC part with both a superfluid and non-supefluid component, which are not separated after 5\unit{ms} TOF.  
However, after 10\unit{ms} of TOF, the three components can be separated. For $T$ just above \TBKT, there is a bimodal distribution (Fig.~\ref{10mstof}, left): the thermal component (solid/blue) and quasi-condensate  (dot-dashed/red). For $T$ just below \TBKT~(Fig.~\ref{10mstof}, right), we see a trimodal distribution:  the thermal component and the non-superfluid QC (which are very similar to the case where $T$ is just above \TBKT) and the superfluid part (very narrow peak, dashed/orange). This transition occurs when the central density reaches $\approx$4.5~atoms/$\mathrm{\mu m}^2$, but with the approximate correction factor to the \textit{in situ} value gives $\approx$9.4 atoms/$\mathrm{\mu m}^2$, which is in reasonable agreement with the theoretical prediction and the 5\unit{ms} TOF analysis.

\begin{figure}
	\center
	\includegraphics[width=.9\linewidth]{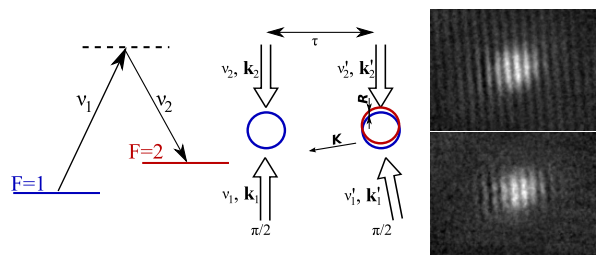}
	\caption{\label{fig:RamanScheme}Left: Schematic of the atom interferometer used to measure the spatial coherence of the 2D atomic cloud. The $\nu$s and $\mathbf{k}$s denote the frequencies and corresponding wavevectors, respectively, of the Raman laser beams. Right: Images of atoms in the $F=2$ state after the interferometer sequence for two different delays between Raman pulses (upper: $\tau=7.5\unit{\mu s}$, lower: $\tau=17.5\unit{\mu s}$). For the longer delay, the interference of the thermal cloud is washed out and there are only fringes from the condensate.}
\end{figure}

To measure the coherence of our gas, we have developed a Ramsey-like method. This allows us to measure coherence properties on a length scale below the thermal de~Broglie wavelength. 
We use two, $\pi/2$, two-photon Raman pulses, with counterpropagating and nearly counterpropagating laser beams, to transfer atoms from the $F=1$ to the $F=2$ internal state, also transferring two photon recoil momenta. We interfere, \emph{in situ}, the two $F=2$ copies of the cloud, which have a velocity different by $\hbar \mathbf{K}/m$ and are spatially shifted by $\mathbf{R}$ (see Fig.~\ref{fig:RamanScheme} and below).
Specifically, if $\hat{\psi}_0(\mathbf{r})$ is the quantum field operator, with
$n_0(\mathbf{r})=
		\braket{
			\hat{\psi}_0^\dagger(\mathbf{r})
			\hat{\psi}_0(\mathbf{r})
				}$
the initial atomic density (in $F=1$), then the measured density of atoms in $F=2$ after the two Raman pulses is 
\begin{equation}
	n(\mathbf{r}) = \frac14\left( 2n_0(\mathbf{r}) + 
	\braket{ 
			\hat{\psi}_0^\dagger(\mathbf{r})
			\hat{\psi}_0(\mathbf{r}-\mathbf{R})
			e^{i\mathbf{r}\cdot\mathbf{K}+i\phi_0}
			+\mathrm{h.c.} 
			}
 \right),
	\label{eq:contrast}
\end{equation}
where $\mathbf{K}=\mathbf{k_1}-\mathbf{k_2}-\mathbf{k^\prime_1}+\mathbf{k^\prime_2}$ (notation defined in Fig.~\ref{fig:RamanScheme}), $\mathbf{R} = \hbar (\mathbf{k_1}-\mathbf{k_2})\tau/m$ is the displacement of the cloud during the time between pulses, $\tau$, 
and $\phi_0$ is the (uncontrolled) phase between the two Raman pulses. By measuring the average spatial contrast of the interference fringes in an image, averaged over several images, we obtain the normalized correlation function 
$g^{(1)}(\mathbf{R}) = \braket{
	\hat{\psi}_0^\dagger(\mathbf{r})
			\hat{\psi}_0(\mathbf{r}-\mathbf{R})
										} / n_0(\mathbf{r})$. 
\begin{figure}
	\includegraphics[width=.9\linewidth]{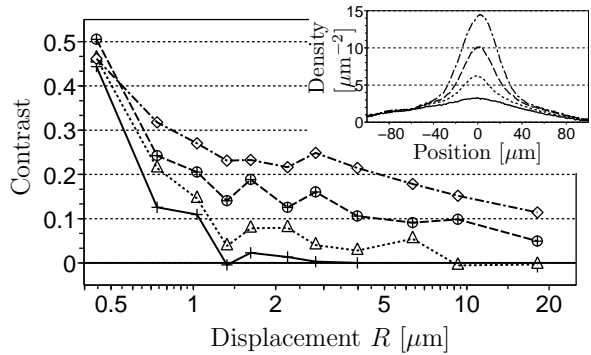}
	\caption{	\label{Fig:ContrastvsDelay}
	Contrast of the interference fringes as a function of the separation of the two clouds in the interferometer. Each curve corresponds to an average over many data sets where the maximum density of atoms falls within a chosen range. Inset: Corresponding density profiles of the atomic cloud. Lines (solid, dotted, dashed, dot-dashed) correspond respectively to density ranges of 2-4, 5-8, 10-12, 13-19 \unit{atoms/\mu m^2}.}
\end{figure}

Figure \ref{Fig:ContrastvsDelay} shows the contrast of the fringes as a function of $\mathbf{R}$, for different densities, but the same temperature, $T\approx 100\unit{nK}$. This contrast is calculated by measuring the relative intensity of the peak of the Fourier component of the image at the spatial frequency of the fringes. Since we cannot precisely control the number of atoms, we average only the data that falls within a chosen range of peak densities. For each curve, we see an initial drop of the contrast on a length scale of the order of 1\unit{\mu m}, due to the relatively short coherence length of the thermal component (the thermal de~Broglie wavelength is $\lambda = 1\unit{\mu m}$ for $T=100\unit{nK}$). For atomic clouds with densities between 5 and 8\unit{atoms/\mu m^{2}} (dotted line), where we clearly see a bimodal distribution, the coherence extends well beyond $\lambda$, but decreasing to zero by $10~\unit{\mu m}$, a distance much shorter than the spatial width of the narrow part of the bimodal distribution (see inset). This behavior is as expected for this quasi-condensate region for $T$ just above $\TBKT$ where, due to local condensation, there is coherence for distances greater than the thermal de~Broglie wavelength, but because of the free vortices, the coherence decreases exponentially on a length scale of the order of $r_c = 2\lambda/\sqrt{2\pi\tilde g} \simeq 6\unit{\mu m}$ \cite{Prokofev}. For higher densities ($T<\TBKT$), the coherence also decays slowly but has a non zero value even at 20\unit{\mu m}. 
For this phase, one expects a power law decay of the coherence mainly due to long wavelength phonons. The typical distance at which the coherence decreases by a factor of 2 is given by $l = 2^{1/\alpha}/\sqrt{\tilde{g}n}$~\cite{Petrov2004}, where $\alpha$ is the coefficient of the power law decay whose value is $1/4$ at the BKT transition. We calculate $l\approx 40\unit{\mu m}$ for our typical parameters  at the BKT transition. 

With our interferometric method, an isolated vortex would be characterized by a double fork structure in the interference pattern \cite{Chevy2001}. To see such a structure, one should separate the two interfering copies of the gas by a distance larger than the fringe spacing ($\approx7\unit{\mu m}$), a distance which is comparable to the healing length ($r_c \simeq6\unit{\mu m}$). However, in the non-superfluid region ($T>\TBKT$), the fringes disappear at this separation. Therefore, with our fringe spacing (constrained by our optical resolution), it is not possible to see such a direct signature of the free vortices that are supposed to proliferate in this regime. This should also holds true in \cite{Hadzibabic2006}, as suggested by the results of \cite{Kruger2007}. 

In this experiment, we have observed the predicted intermediate regime, between the thermal and superfluid phase, characterized by a bimodal density distribution and a coherence length longer than the thermal de~Broglie wavelength, but smaller than the predicted length scale in the superfluid region. 
At higher densities, we see a phase with a much longer coherence length, which is consistent with the BKT superfluid phase. For sufficiently long TOF we clearly see a trimodal distribution showing the presence of three components. A possible explanation for previous experiments \cite{Hadzibabic2006, Kruger2007} having missed seeing a trimodal distribution (they associate the BKT transition with the appearance of bimodality) is that, they don't see this intermediate quasi-condensed regime. This may be because their imaging integrates over one dimension, reducing the signal of the narrow part of the distribution relative to that of the thermal cloud. Finally, our expermental setup, which allows us to see BKT related physics in a single "pancake" of atoms, opens new oportunities, such as studying density fluctuations at the transition. Future experiments could use a Feshbach resonance to tune the coupling constant in our all optical trap, and explore different interaction regimes.

This work was suported by the Office of Naval Research. We thank J.~Dalibard, T.~Simula, M.~Davis and B.~Blakie for helpful discussions.


\begin{thebibliography}{20}
\bibitem{uncertainties}{Unless otherwise stated, the uncertainties are one standard deviation combined systematic and statistical.}
\end{thebibliography}
\end{document}